\documentclass[a4paper]{jpconf}
\usepackage{graphicx}
\begin{document}
\title{Thermal stability of standalone silicene sheet}

\author{Virgile Bocchetti$^{1,a}$, H. T. Diep$^{1,b,}$\footnote[5]{Corresponding author}, Hanna Enriquez$^{2,c}$, Hamid Oughaddou$^{3,d}$ and Addelkader Kara$^{4,e}$}

\address{$^{1}$  Laboratoire de Physique Th\'eorique et Mod\'elisation,
Universit\'e de Cergy-Pontoise, CNRS, UMR 8089,
2, Avenue Adolphe Chauvin, 95302 Cergy-Pontoise Cedex, France}

\address{$^{2}$ Institut des Sciences Mol\'eculaires d'Orsay, ISMO-CNRS, B\^atiment 210, Universit\'e Paris-Sud, 91405 Orsay, France}

\address{$^{3}$ D\'epartement de Physique, B\^atiment Neuville II, 5 Mail Gay-Lussac, 95031 Neuville sur Oise, Cergy-Pontoise Cedex,
France}

\address{$^{4}$ Department of Physics, University of Central Florida, Orlando, Florida, 32816 USA}

\ead{$^{a}$virgile.bocchetti@u-cergy.fr, $^{b}$diep@u-cergy.fr,
$^{c}$hanna.enriquez@u-psud.fr, $^{d}$hamid.oughaddou@u-cergy.fr,
$^{e}$Abdelkader.Kara@ucf.edu}

\begin{abstract}
Extensive Monte Carlo simulations are carried out to study thermal stability of an infinite standalone silicon sheet.
We used the Tersoff potential that has been used with success for silicon  at low temperatures.  However, the melting temperature $T_m$ calculated with the original parameters provided by Tersoff is too high with respect to the experimental one. Agrawal, Raff and Komanduri  have proposed a modified set of parameters to reduce $T_m$. For comparison, we have used these two sets of parameters
to study the stability and the melting of a standalone 2D sheet of silicon called "silicene",
by analogy with graphene for the carbon sheet.
We find that 2D crystalline silicene is stable up to a high temperature unlike in 2D systems with isotropic potentials such as Lennard-Jones.
The differences in the obtained results using two sets of parameters are striking.
\end{abstract}

\section{\label{sec:level1}Introduction}
Physics of 2D systems has been spectacularly developed during the last 30 years both experimentally and theoretically
due to numerous applications in nanotechnology.  In particular, we can mention the recent discovery of graphene with unusual
properties such as high stability of 2D structure and remarkable transport behavior.
For recent reviews given by experts on fundamentals and functionalities of graphene, the reader is referred to a special volume of MRS Bulletin \cite{MRS2012}. Graphene is a Carbon sheet of one-atom layer thickness with large lateral size (up to a dozen of micrometers) discovered in 2004 \cite{Novoselov,Berger}. It has striking properties which can be applied in many domains such as optical materials, electronic materials, terahertz technology and spintronics.
Intensive research activities are being carried out to incorporate graphene into devices for applications to exploit
the novel properties of this unique nanomaterial (see recent reviews in Refs. \cite{Singh,Haas,Allen,Sarma,Fuhrer,Araujo,Kuilla}
This is not
an easy task. This difficulty leads scientists to turn to an alternative 2D version of silicon or germanium which would be easier to incorporate into the current silicon-based technology. In this paper we
focus our attention in 2D silicon sheet known as silicene.
Experiments have shown the existence of silicene nanoribbons \cite{Kara2012,Aufray2010} on Ag(110) surface and a silicene sheet on Ag(111) surface \cite{Lalmi2010,Lin2012,Jamgoitcha2012,Feng2012,Gao}. An ordered two dimensional surface alloy has been observed upon adsorption of silicon on Au(110) \cite{Enriquez2012} and on a ZrB2 substrate \cite{Fleurence}.
Theoretical works using Molecular-Dynamics (MD) simulations \cite{Ince2011} and density-functional theory (DFT) \cite{Kara2010} have shown the
stability of nanoribbons of different widths against temperature. Evidence of graphene-like electronic signature in silicene nanoribbons has been shown \cite{Padova2010} and electronic structures of silicene fluoride and hydride have been revealed by calculations \cite{Ding}.
There is also a number of investigations on the stability of silicene sheet  using
a structure optimization \cite{Cahangirov2009}, phonon dispersion and \textit{ab initio} finite-temperature DFT theory.  A low buckling of the silicene sheet has been suggested in these works.
One of the problems in dealing with a silicon crystal concerns the choice of an appropriate interaction potential between Si atoms. In the bulk,  a few potentials have been proposed to describe properties of bulk silicon crystals such as the Stillinger-Weber \cite{Stillinger1985} and Tersoff potentials \cite{Review1989,Letters1986}.  These potentials have succeeded to reproduce principal physical properties of bulk Si crystal at low temperatures and to stabilize the diamond structure  up to
very high temperatures. However as seen below the melting temperature is largely over estimated with those potentials using MD simulations.  To our knowledge, no Monte Carlo simulations on the subject have been reported, which triggered the interest in the subject..

In this paper, we show our results of constant-pressure MC simulations on the stability of a  silicene sheet.
In  section \ref{sec:level2} we present the Tersoff potential, our method and algorithm. In section \ref{sec:level3} we present our results  obtained with potentials using two different sets of parameters.

\section{\label{sec:level2}Model and Monte Carlo method}
\vspace{0.5cm}
\subsection{Tersoff potential}
One of the most used potentials with the Metropolis MC algorithm is the Lennard-Jones potential (LJ) \cite{Bernardes1958}.
This potential cannot be employed to describe the silicene honeycomb structure. Such an isotropic potential always leads the system into the most compact face-centered-cubic (FCC)
lattice when the temperature decreases. For Si crystals, the two potentials which are suitable at least at low temperatures are the Stillinger-Weber and Tersoff potentials. In this paper, we use the Tersoff potential. The reason of this choice will be explained below.
This empirical potential was introduced by Tersoff in order to improve the accuracy in the description of properties of the Si crystal. Tersoff potential takes into account the dependence of the bond configuration on the local symmetry. This potential was introduced after the work of Ferrante, Smith, and Rose who
have shown the universal behavior of calculated binding-energy curves for solid cohesion \cite{Review1983,Rose1983}.
We used here the Tersoff  potential \cite{Review1989} given by the following
expression:


\begin{equation}
V_{Tersof} = \frac{1}{2} \*{\sum_ {{i \ne j}}{V_{ij}}}
\end{equation}
where
\begin{equation}
V_{ij} = f_c(r_{ij})\left[A\exp\left( -\lambda r_{ij}\right)-b_{ij}B\exp\left( -\mu r_{ij}\right)\right]
\end{equation}
\begin{equation}
b_{ij}=\chi \left(1 + \beta^n \zeta_{ij}^n \right)^{\frac{-1}{2n}}
\end{equation}
\begin{equation}
\zeta_{ij} = \sum_{k \ne i,j}f_c(r_{ik})g\left( \theta_{ijk}\right)
\end{equation}
where
\begin{equation}
g\left( \theta_{ijk}\right) = 1+\frac{c^2}{d^2}-\frac{c^2}{\left[ d^2+\left(h-\cos \theta_{ijk}\right)^2\right]}
\end{equation}
and the cut-off function
\begin{equation}
f_c(r_{ij}) = \left \{%
\begin{array}{ll}
1 &  \mbox{if}\    r_{ij} < R,\\
\frac{1}{2} + \frac{1}{2} \cos \left[\pi \frac{\left( r_{ij}-R\right)}{S-R}\right] &  \mbox{if}\  R < r_{ij} < S,\\
0 &  \mbox{if}\  r_{ij} > S \\
\end{array} \right \}%
\end{equation}
In the above expressions, $A$, $B$, $c$, $d$, $h$, $n$, $\lambda$, $\mu$, $\beta$, $\chi$, $R$ and $S$ are constants.
The above potential has two terms: the $A$ term and the $B$ term (see expression of $V_{ij}$). The $A$ term is the two-body interaction while
the $B$ term incorporates many-body interactions including the three-body angles $\theta_{ijk}$.   This term determines the
 diamond structure of 3D silicon. Looking closely at the structure of the Tersoff potential, we realize that the two most important
parameters are $n$ which fixes the many-body interaction strength and $h$ the parameter which determines the angle between two connected bonds.
The original parameters given by Tersoff \cite{Review1989,Letters1986} which yield a too high melting transition temperature $(T_m=3600$ K) have been modified by Agrawal {\it et al.} to reduce it down to 2200 K which is closer to the experimental value $T_m$(exp)$\simeq 1690$ K.
These two different sets of parameters for the Tersoff potential are given in Table \ref{tab:table1}.  Note that the main modifications concern precisely $n$ and $h$, as discussed above.
\begin{center}
\begin{table}[h]
\caption{\label{tab:table1}%
Tersoff parameters}
\centering
\begin{tabular}{@{}*{7}{l}}
\br
Parameter&Orginal values (Ref. \cite{Review1989})
&Values from ARK (Ref. \cite{Agrawal2005})\\
\mr
A \ (eV) & 1830.8 & 1830.8\\
B \ (eV) & 471.18& 471.18\\
$\lambda$ \  (\AA$^{-1}$) & 2.4799& 2.4799\\
$\mu  $ \  (\AA$^{-1}$) & 1.7322& 1.7322\\
$\beta$ & $1.1\times 10^{-6}$& \textbf{1.15}$\times$\textbf{10}$^{-6}$ \\
n & 0.78734 &\textbf{0.988} \\
c & $1.0039\times10^5$&  $1.0039\times10^5$\\
d & 16.217& 16.217\\
h & -0.59825 & \textbf{-0.74525}\\
R \ (\AA) & 2.7& 2.7 \\
S \ (\AA) & 3.0& 3.0\\
$\chi$ & 1.0& 1.0\\
$T_m$ (Bulk Si) & 3600\ {\mbox{K}}& 2200 \ {\mbox{K}}\\
\br
\end{tabular}
\end{table}
\end{center}

\subsection{The Monte Carlo algorithm}
In our simulations, we consider a system of 968 atoms.
The algorithm is split into two main parts: the construction of the honeycomb lattice with the minimization of the lattice energy at 0 Kelvin, and
in the second part, the MC algorithm using the Metropolis updating criterion \cite{Allen1987}.

\begin{figure}[h]
\begin{center}
\includegraphics[width=4.3cm,height=5cm]{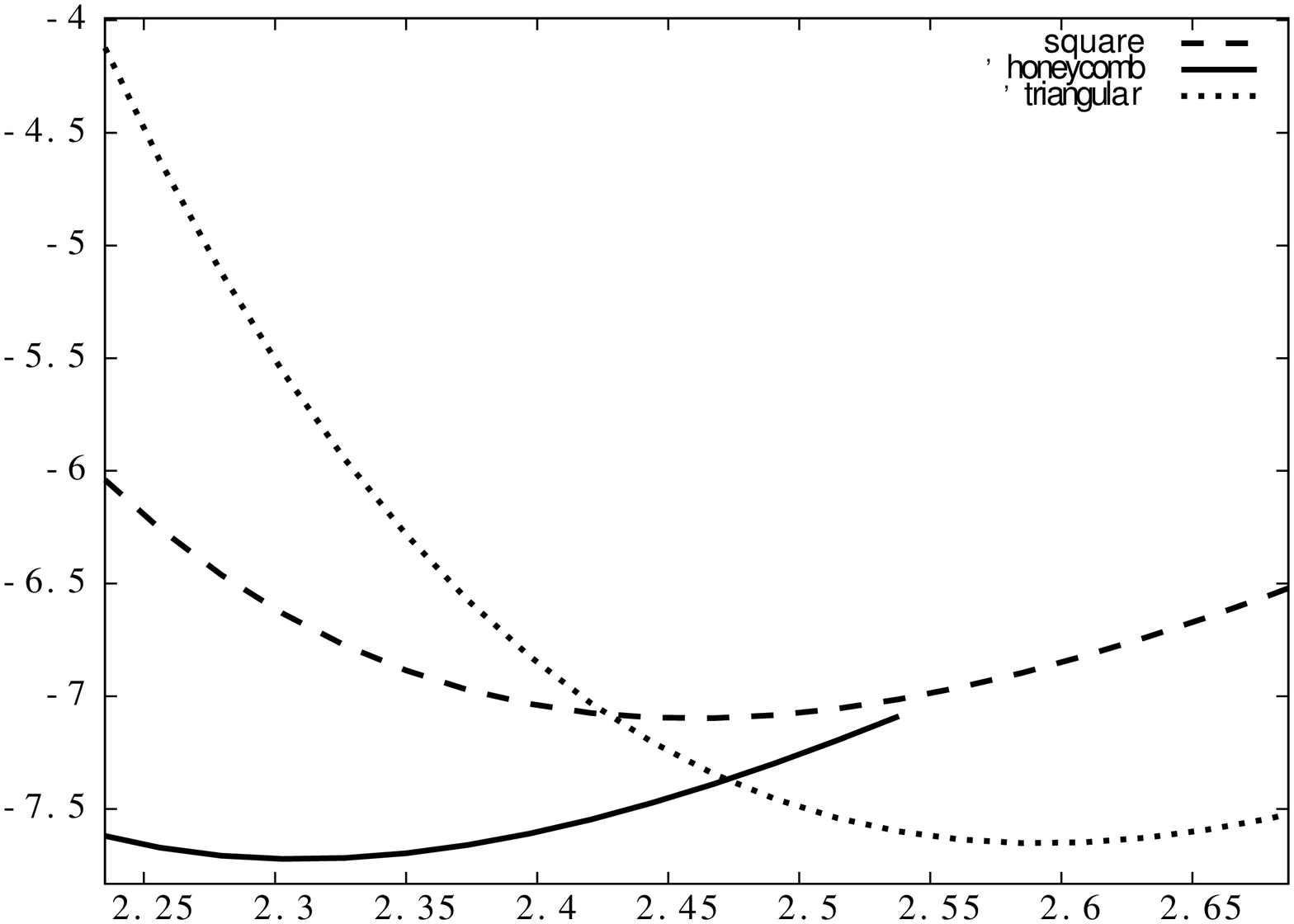}
\includegraphics[width=4.3cm,height=5cm]{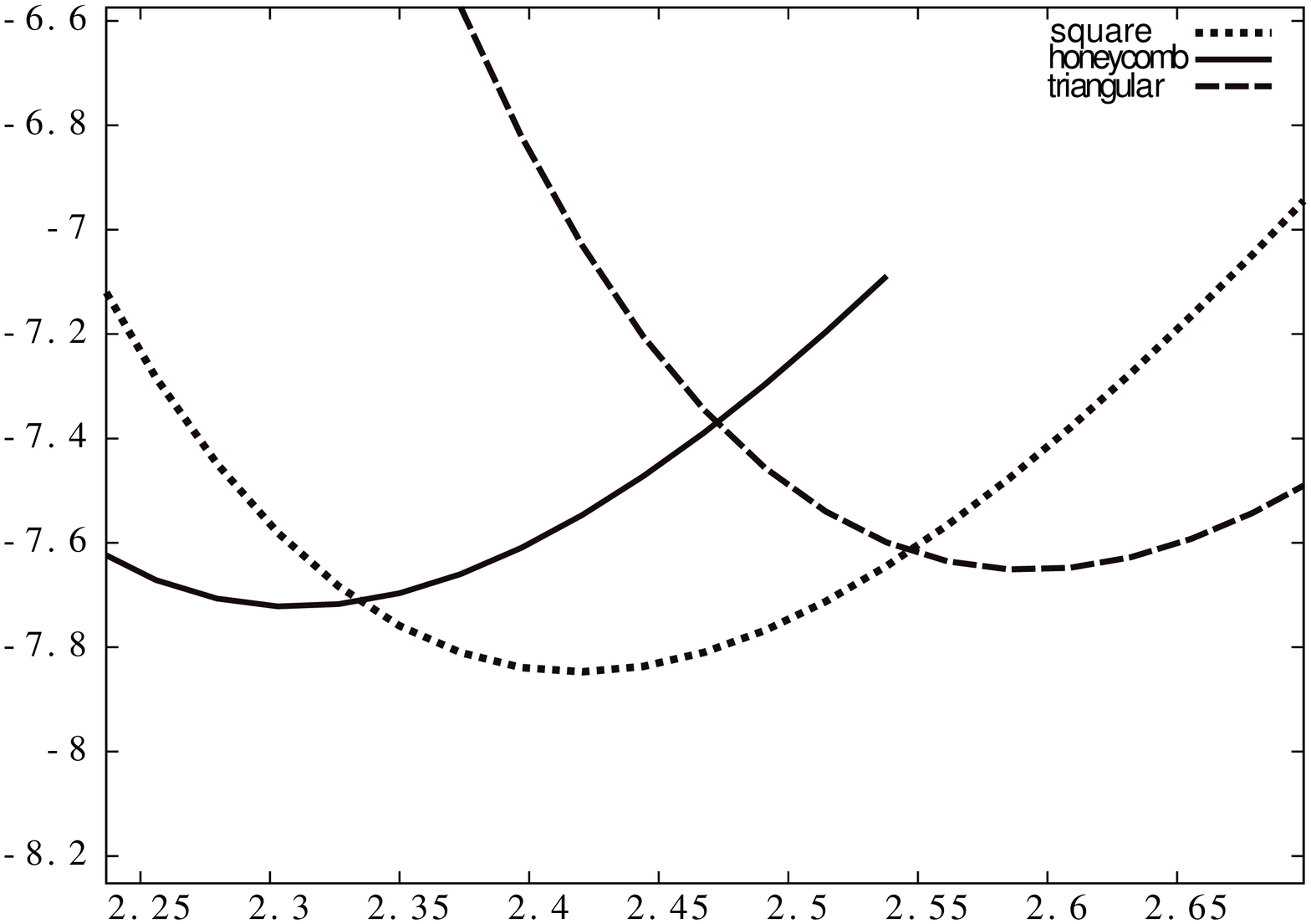}
\end{center}
\caption{\label{reso} Energy per atom at 0 Kelvin,for three different lattices with
 the original set of parameters (left) and the ARK set of parameters (right). See text for comments.}
\end{figure}

For testing purpose, we have build three different 2D  planar lattices
 of silicon, namely honeycomb, square and triangular structures, and we have computed the energy per atom in this three different configurations. The results are shown in Fig. \ref{reso}.
As we can see, the honeycomb lattice is more stable than the two others at the nearest-neighbor (NN) distance $r=2.31$ \AA which agrees with experiments, using both the original set of parameters given by Tersoff \cite{Review1989} and the ARK parameters \cite{Agrawal2005}: the energy has  a minimum for the honeycomb lattice at that NN distance.
 For larger NN distances, the ARK parameters give an energy minimum for the square lattice at $r =2.42$ \AA, but this distance does not correspond to the NN distance between  Si atoms in the silicon crystal.  So, at $T=0$, we can say that both potentials  give the same energy $E_0\simeq 7.7 $ eV and the same NN distance $2.31$ \AA.



Furthermore, calculations based on DFT \cite{Cahangirov2009} found that a buckling in the honeycomb structure of 0.4 {\AA} stabilizes silicene.
 The buckled configuration is the stacking
 of two inverted triangular planes at a small $z$ distance.
In order to check this suggestion with our potential,
we have introduced a parameter $d$
which is the distance between two triangular planes and we have computed the energy per atom as a function of $d$.
The potential used in the present study gives the planar structure ($d=0$) as the energetically favorable structure. In order to reproduced the buckling found by the more accurate calculations, one need to re-fit all the parameters of the potential with the addition of one (such as $d$) ore more parameters; this is out of the scope of this study and will be tackled in futures studies.


For each MC  step (MCS) we move all the atoms and we relax the size of the system. The magnitudes of the atomic displacements and the variation of the system size  are determined so as to obtain an acceptation rate  of  about $50\%$.
This collective updating is different from the single-atom updating algorithm which is not at all efficient for melting studies.
Our algorithm allows a variation ratio of the volume; so the volume can make fluctuations (dilatation and contraction) around its equilibrium during the simulation time.
An important fact is that the volume variation is controlled by the Metropolis algorithm, like the atom positions.
For each simulation, we made approximately $3.10^7\ $ MC steps per atom.  Such long runs allow to observe the stability of the system and to overcome the very long relaxation time near the melting.
At each MC step, after updating atom positions and relaxing the system volume, we compute the following transition probability

\begin{equation}
W = P\  \left(V_{new} - V_{old} \right) +  \left(U_{new} - U_{old}  \right) +
N \ k_{B} \ T \ \ln\left({\frac{S_{new}}{S_{old}}}\right)
\end{equation}
where $P$ is the pressure ( $0$ in our case ), $S_{new}$ and $S_{old}$ are respectively the new system surface and the old one, $U_{new}$ denotes the new energy of  the system after trial updating,
$U_{old}$ the old one, $k_{B}$ the Boltzmann constant and $T$ the temperature ( in Kelvin ).

A trial move is accepted if
\begin{equation}
\xi \le  e^{\left(\frac{-W}{k_{B}T} \right)}
\end{equation}
where  $\xi$ is a random number between $0$ and $1$. Otherwise, the system  returns to its  previous state with old atom positions and old surface size.  We tune the magnitude of atom displacements and volume variations so as to have an acceptation rate of around $50\%$.

\section{\label{sec:level3}Results and discussion}

For 2D systems with short-range isotropic interaction, it is known that long-range order does not survive at finite temperatures \cite{Mermin1968,Nelson1979}.  Melting transition at finite $T$ predicted by the Linderman criterion is for 3D crystals \cite{Linderman1910}.
For the present 2D silicene, the potential is not isotropic because it stabilizes the Si diamond structure at very high $T$. As it turns out, this potential stabilizes also the honeycomb structure, as seen below.
The stability of a silicene sheet can be observed by the energy versus temperature curve, the radial distribution function, snapshots of the system, the
angular distribution function  or the structure factor. We will show these quantities below.

In all our simulations, we started with a perfect lattice at 0 K and we increase the temperature to the interested temperature range.

In order to have more independent data and also to have faster computation, for each temperature we  compute  physical quantities of the system
 on an independent node of  a CPU cluster. As an example, we show in
Fig.\ref{ETERP} the mean energy against temperature where each data point was computed by a node.

\subsection{Results using original Tersoff parameters}
\begin{figure}
\begin{center}
\includegraphics[width=8cm,height=5cm]{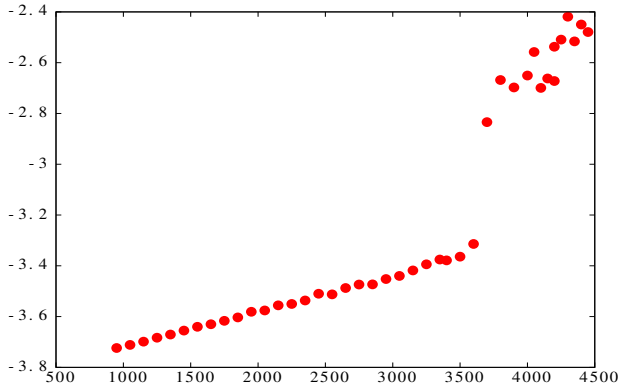}
\end{center}
\caption{\label{ETERP}Energy versus temperature using original Tersoff parameters.}
\end{figure}
\begin{figure}
\begin{center}
\includegraphics[width=8cm,height=5cm]{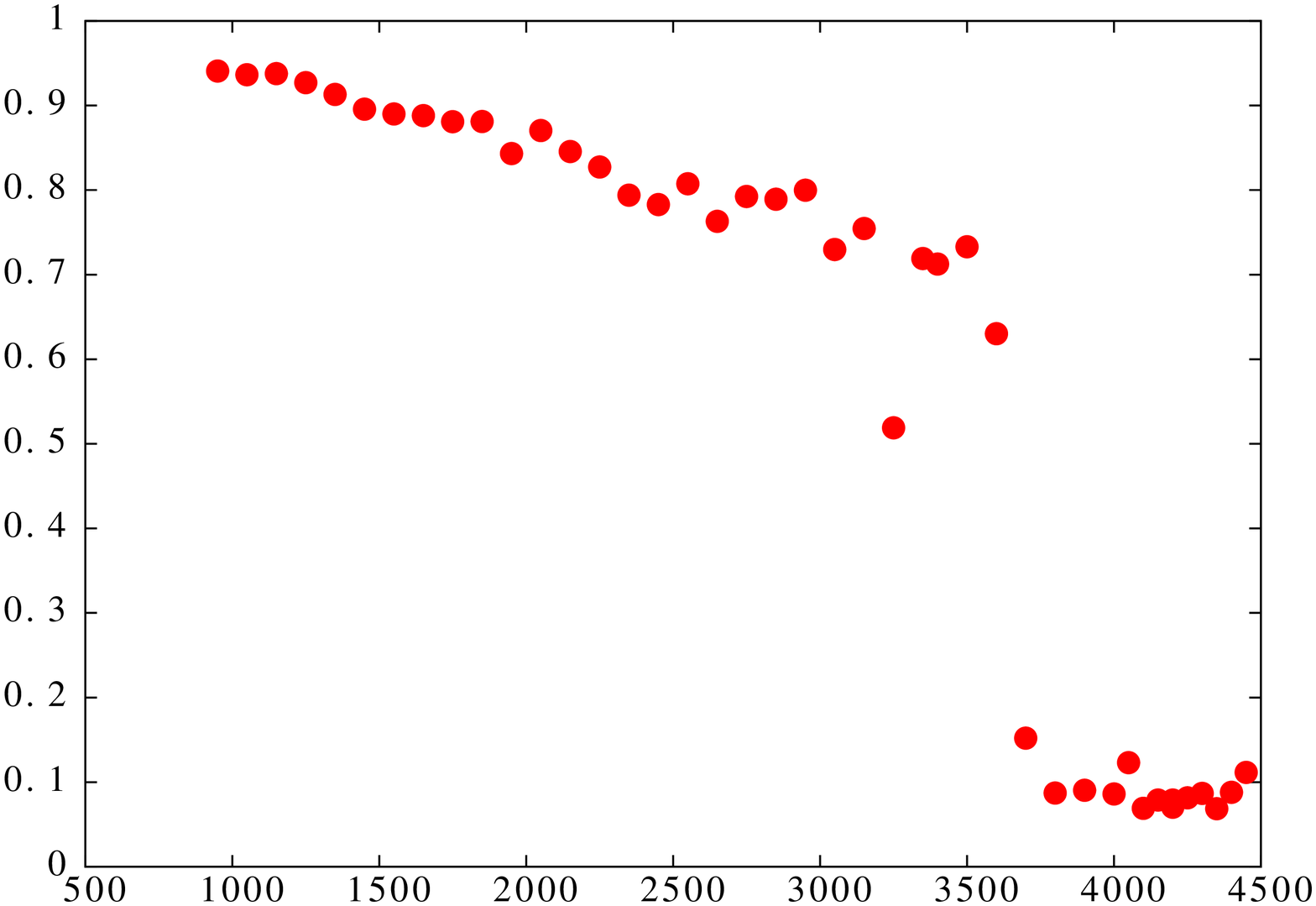}
\end{center}
\caption{\label{f_s_TER}Structure factor versus temperature using original Tersoff parameters.}
\end{figure}

At high $T$ (see Fig. \ref{ETERP}), we observe a first-order transition with a large latent heat.  This transition is the melting of the sheet.
Note that the melting temperature is very close to the simulated melting temperature of a 3D silicon crystal ($\simeq 3600$ K) using the same potential.
We calculate the structure factor  $S_{\vec K}$  defined as

\begin{equation}
S_{\vec K} =\frac{1}{N_l}\left < \left |\sum_{j=1}^{N_l}e^{i\overrightarrow{K}.\overrightarrow{d_j}}\right |\right >
\end{equation}
where $\overrightarrow{d_j}$ is the position vector of an atom in the layer, $N_l$  the number of atoms in a layer and $\overrightarrow{K}$  the reciprocal lattice vector which has the following coordinate (in reduced units):
$2\pi\left(1;-\sqrt{3};0\right)$.  The angular brackets $<...>$ indicate thermal average taken over MC run time.
The above  "order parameter" allows us to monitor the long-range order. We show in Fig. \ref{f_s_TER} the structure factor versus $T$.
As seen, the long-range order is lost at $T\simeq 3600$ K, namely at the temperature where the energy has a large discontinuity.

\begin{figure}
\begin{center}
\includegraphics[width=8cm,height=5cm]{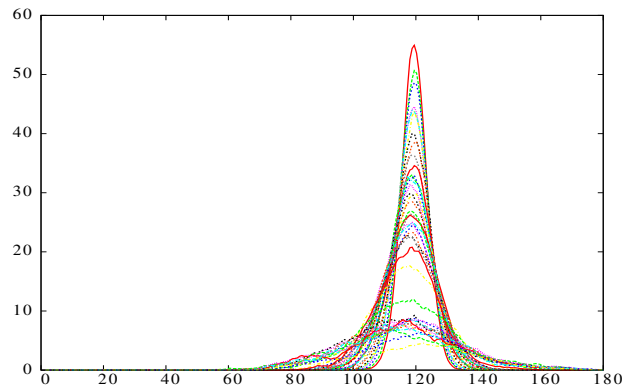}
\end{center}
\caption{\label{adf_TERP}Angular distribution function for different $T$. See text for comments.}
\end{figure}

We show in Fig. \ref{adf_TERP} the angle distribution function at various temperatures.  The pronounced peak at $T< 3500$ K undergoes a discontinuous decrease at $T\simeq 3600$ K.
Furthermore, the angle distribution function does not show the appearance of any peak different from that at $120^\circ$.  This shows the high stability of the honeycomb structure up to melting.

\begin{figure}[h]
\begin{center}
\includegraphics[width=4.5cm,height=5cm]{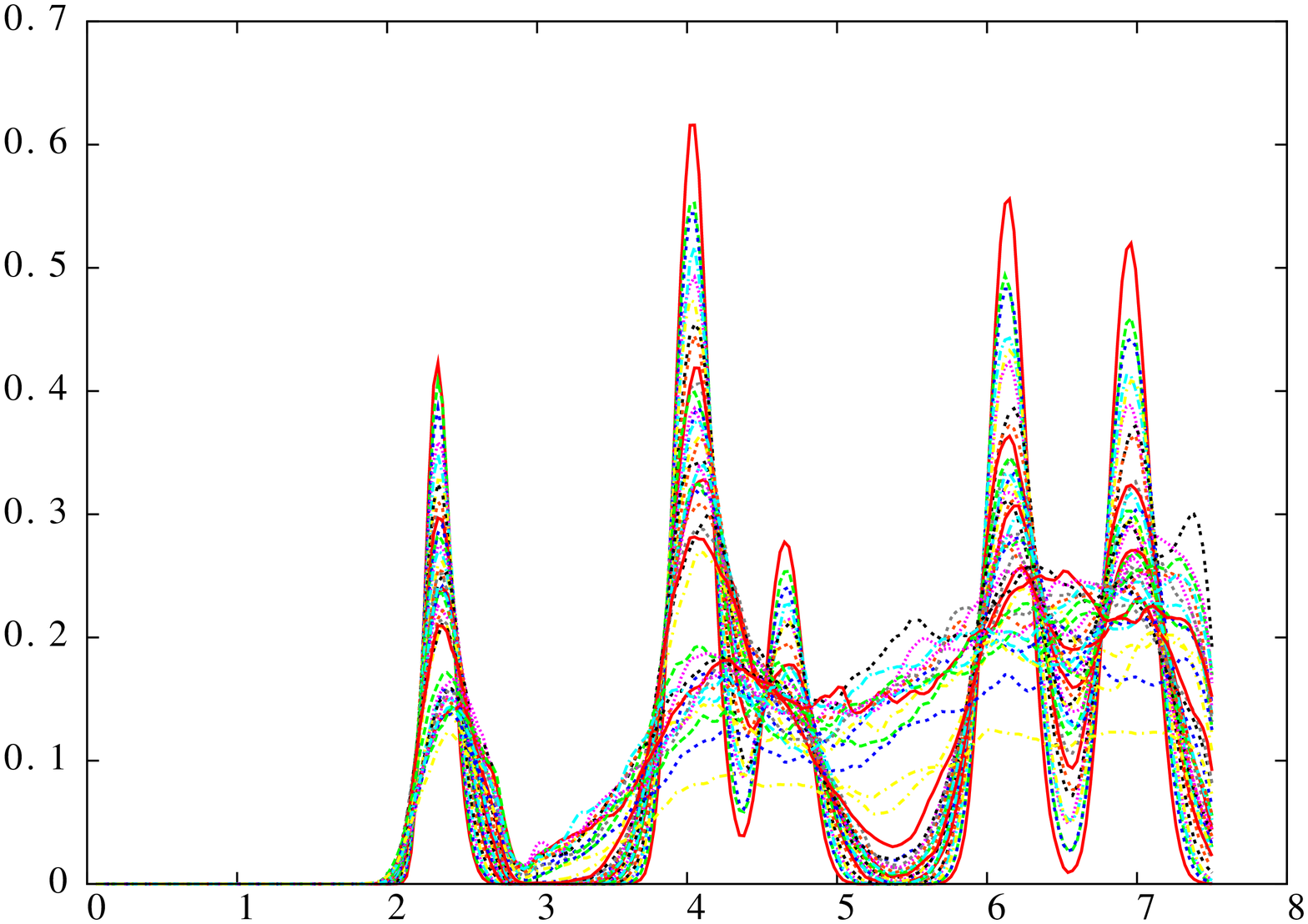}
\includegraphics[width=4.5cm,height=5cm]{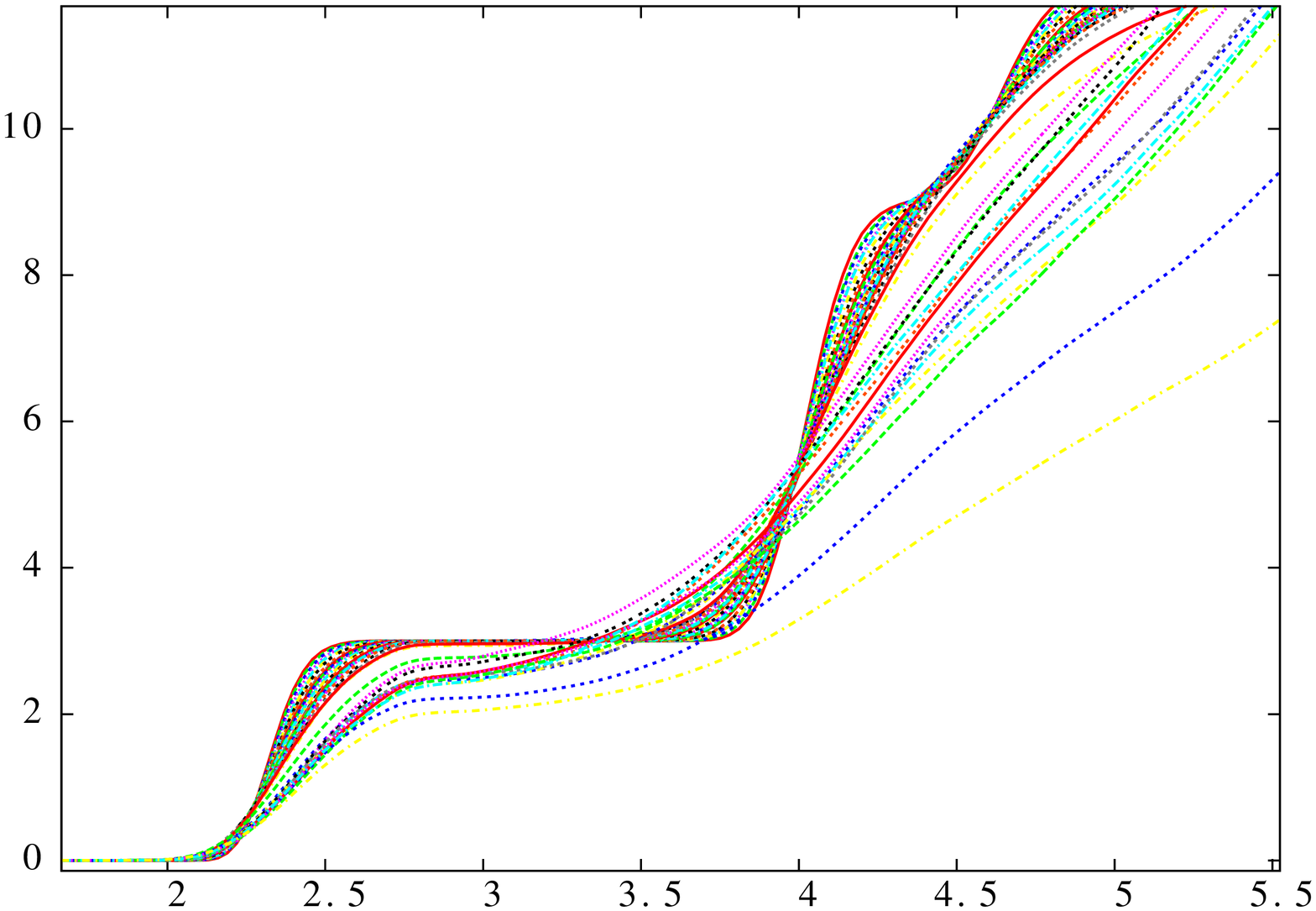}
\end{center}
\caption{\label{integral_TER} Radial distribution function for several $T$ (left).
Integrated radial distribution function for several $T$ (right).  See text for comments.}
\end{figure}

The radial distribution functions confirm the transition.   When the temperature increases, the  radial distribution function (Fig. \ref{integral_TER}) jumps from a state where we can distinguish the peaks corresponding to ordered positions of far neighbors to
a state where only the peak of nearest-neighbors remains. The long-range order is lost.

\begin{figure}
\begin{center}
\includegraphics[width=5cm,height=5cm]{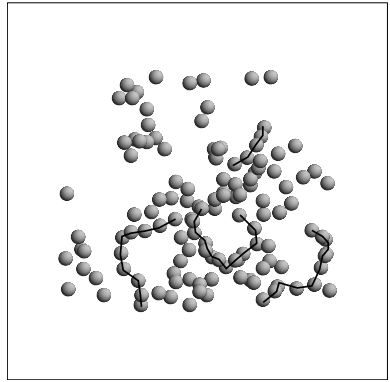}
\end{center}
\caption{\label{wire_TER}Picture of the system at high $T$ (above the transition temperature).  Disordered wire structures are observed.}
\end{figure}

In order to see how the number of nearest-neighbors evolves and indirectly how the density is modified, we have computed the integrated radial distribution shown  in Fig. \ref{integral_TER}.
As we can see, this number decreases from $3$ in the perfect crystal to $2$. The latter density corresponds to a wire structure of Si in 3D space (see Fig. \ref{wire_TER}). This behavior has been observed in the study of the melting of graphene \cite{Zakharchenko2011}.

\subsection{ARK parameters}

\begin{figure}
\begin{center}
\includegraphics[width=8cm,height=5cm]{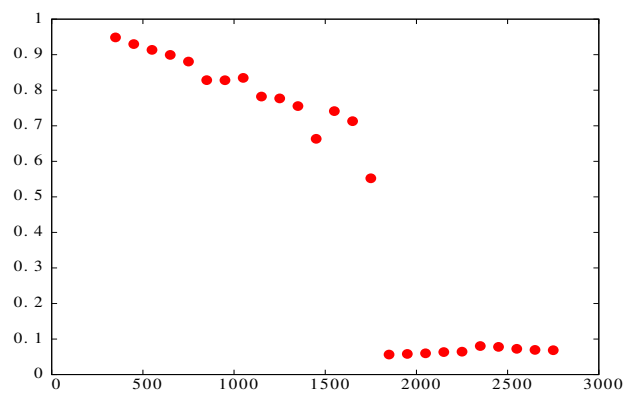}
\end{center}
\caption{\label{f_s_ARK}Structure factor versus temperature with ARK parameters.}
\end{figure}

Let us recall that the experimental value of the bulk melting temperature $T_m$(exp) is about 1700 K. The Tersoff parameters yield $T_m$(Tersoff)$\simeq 3600$ K  while the ARK parameters used for the bulk Si crystal gives $T_m$(ARK)$\simeq 2200$ K.  So,  the ARK parameters give a melting temperature closer to the experimental value.

In the case of a standalone sheet, the original Tersoff parameters, as shown above, give a very high melting temperature, almost identical to that of the bulk, namely $T_m\simeq 3600 $ K.
Let us show now the melting temperature of a standalone sheet obtained by using the ARK parameters. We show  the structure factor in Fig. \ref{f_s_ARK}. The long-range order is lost at $T_m \simeq 1750$ K, lower than that of the ARK bulk value, and less than a half of that obtained by using the original Tersoff parameters..

\begin{figure}
\begin{center}
\includegraphics[width=8cm,height=5cm]{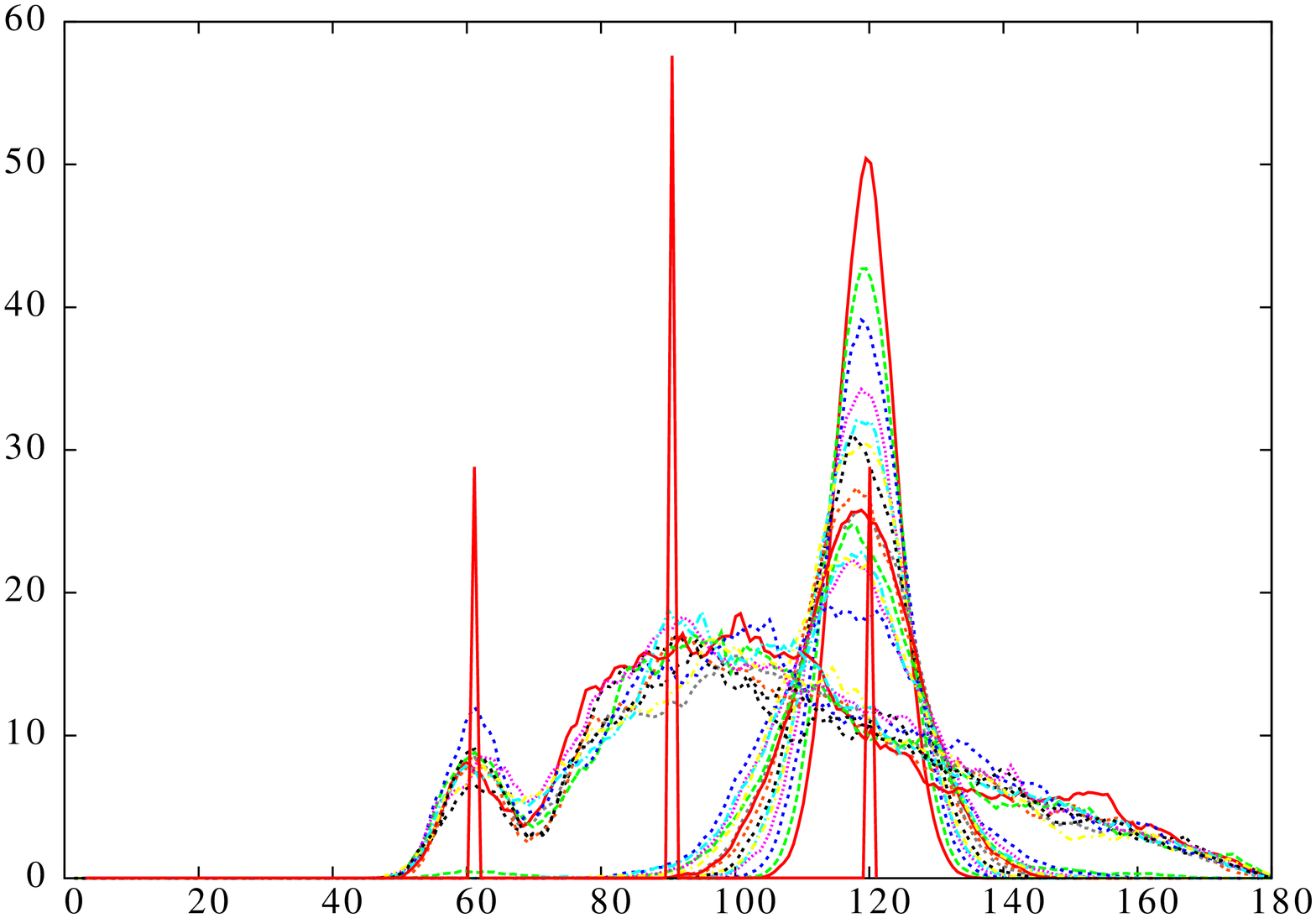}
\end{center}
\caption{\label{adf_ARK} Angular distribution function for a wide range of $T$.
Angular distribution function for the hexagonal 3D structure with three red peaks at $60^\circ$, $90^\circ$ and $120^\circ$ is shown for comparison.}
\end{figure}

 Furthermore, the angular distribution function (see Fig. \ref{adf_ARK}) shows the apparition of two peaks at $60^\circ $ and $90^\circ$ at the same time a decrease of the peak at $120^\circ$.  The honeycomb structure is thus strongly deformed to give rise to a 3D structure. For comparison, we show in Fig. \ref{adf_ARK} the angle distribution of the 3D hexagonal lattice.
This result confirms  the structural transition of the sheet to a 3D film of silicon in a structure where angles of  $60^\circ$ and $90^\circ$ are proliferated.

\begin{figure}[h]
\begin{center}
\includegraphics[width=5cm,height=4cm]{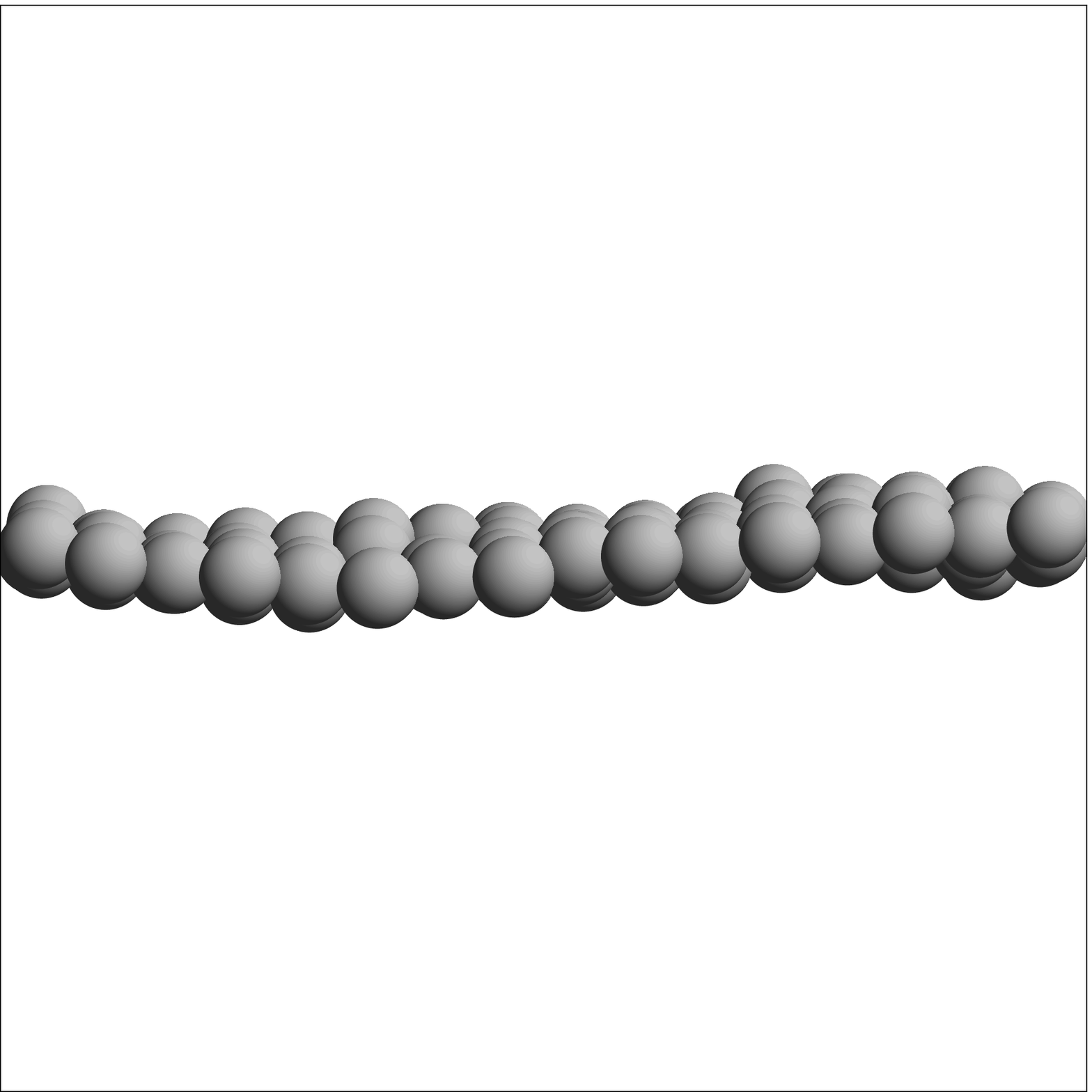}
\includegraphics[width=5cm,height=4cm]{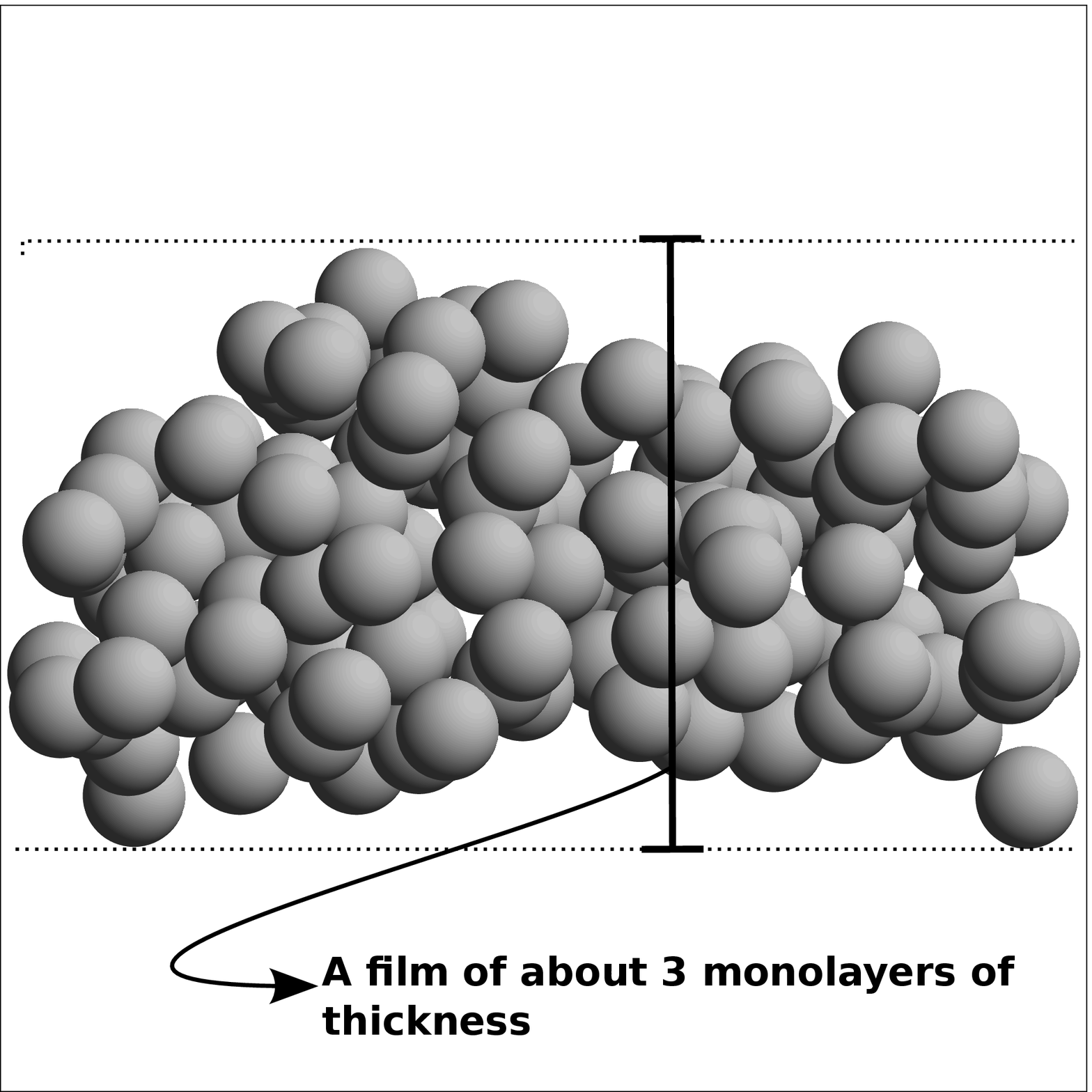}
\end{center}
\caption{\label{image_ARK}Instantaneous snapshots of the system shown by side views:
the 2D structure below the transition (left) becomes a 3D one (right) above the transition.}
\end{figure}

In Fig. \ref{image_ARK}  we show the system at low and high $T$ where we see the 2D structure before the transition and a 3D one at high $T$.

\begin{figure}[h]
\begin{center}
\includegraphics[width=5cm,height=5cm]{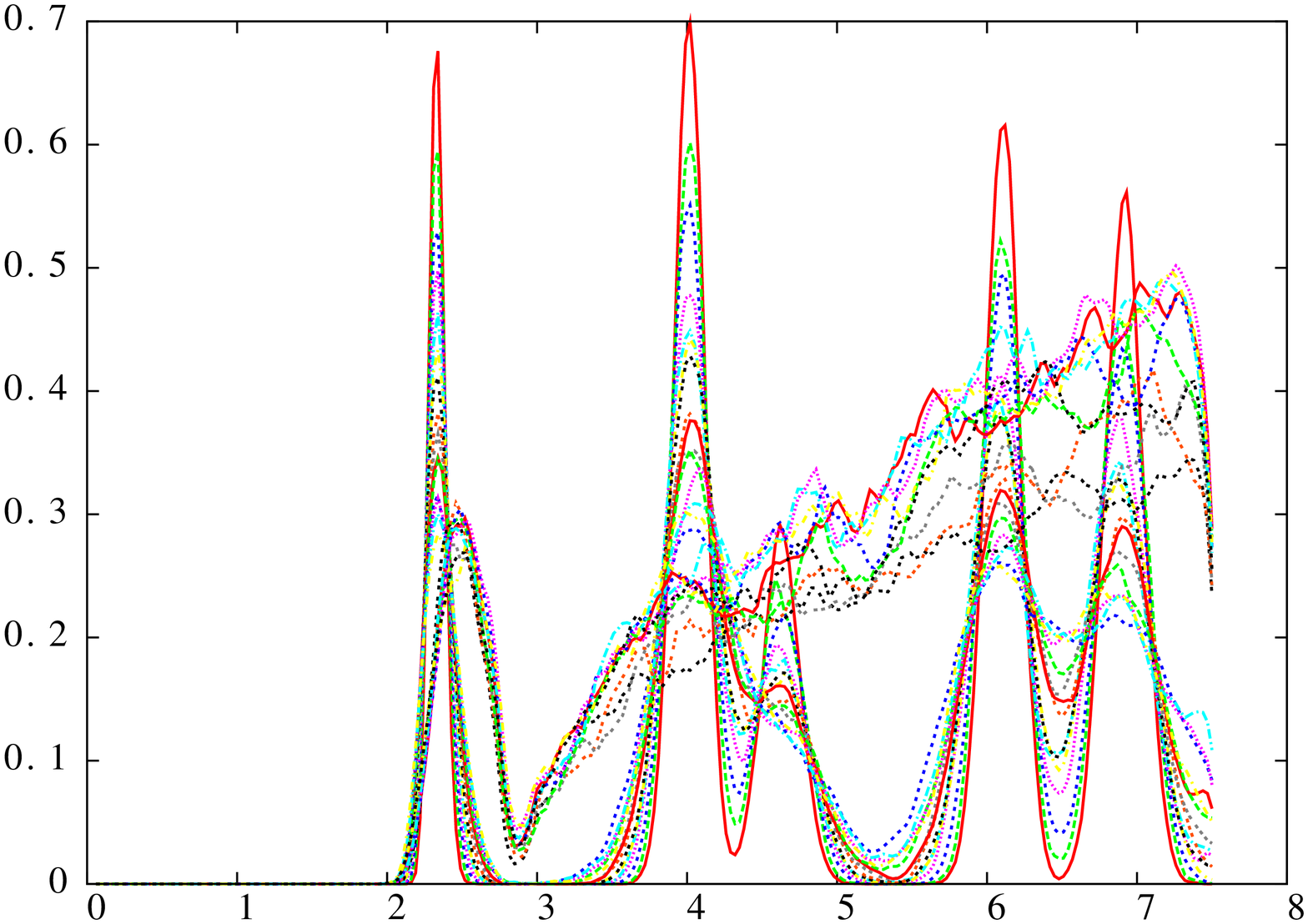}
\includegraphics[width=5cm,height=5cm]{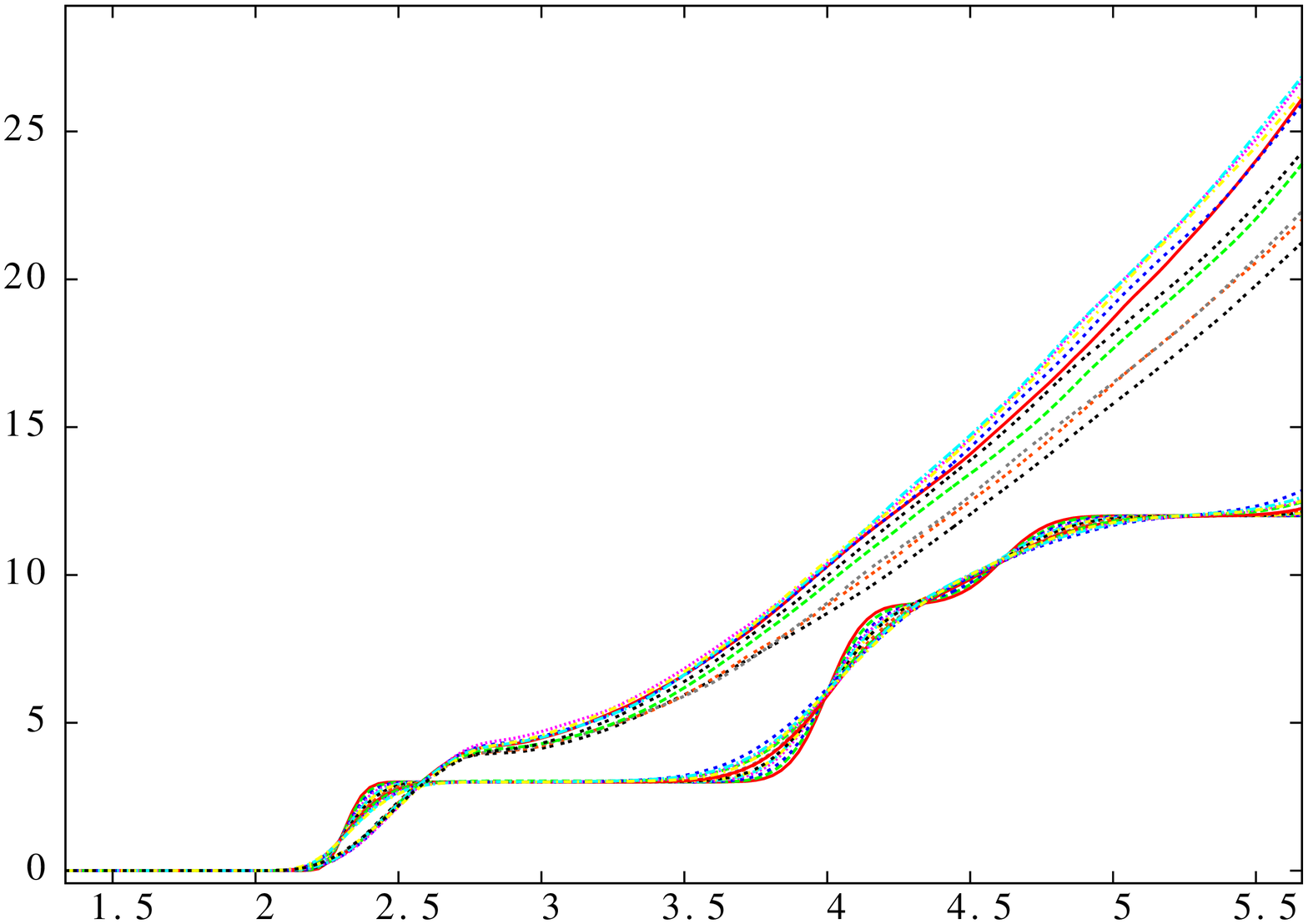}
\end{center}
\caption{\label{integral_ARK}Radial distribution function for different $T$ (left).
Integrated radial distribution function for different $T$ (right).}
\end{figure}

The radial distribution  and the integrated radial distribution shown in Fig. \ref{integral_ARK} indicate an abrupt transition with a jump of the number of nearest neighbors from $3$ at low $T$ to $4.6$ at high $T$. This means that the silicon sheet is reorganized in a 3D structure which is more dense than the previous  2D one. The density jump in the integrated radial distribution is in agreement with the visual observation shown in Fig. \ref{image_ARK}.

\section{\label{sec:level4}Concluding remarks}
In this paper, we studied the behavior of an infinite standalone silicene sheet. We have shown that the 2D honeycomb structure is stable up to high temperatures with the Tersoff potential. However, the temperature range of the silicene stability depends strongly on the  parameters of that potential. The original Tersoff parameters give a too-high melting temperature while those proposed by Agrawal, Raff and Komanduri yield a melting temperature which is a half lower and much closer to 3D Si melting temperature.  For both sets of parameters, the Tersoff potential gives rise to a silicene sheet without  buckling. The flatness is stable with increasing temperature.  Note that a very small buckling has been experimentally observed in the silicon deposition onto Ag(111) substrate
maintained during the growth at 250 C \cite{Lalmi2010} as well as  in a theoretical DFT study \cite{Cahangirov2009}.  There may be several reasons to explain the difference between these works and ours:

i) Bucking depends on the potential: for example with the Stillinger-Weber potential \cite{Stillinger1985} we find that the ground state of silicene is buckled. However, with this potential, the flatness is progressively destroyed even far below the melting temperature.  As said above, if we wish to
introduce a buckling with the Tersoff potential, we should modify the whole set of parameters, not just add $d$ to the potential.

ii) Buckling experimentally observed may come from the interaction with the substrate. Preliminary MC results from a Si ribbon deposited on Ag(111) show that the Si atoms do not remain on a flat plane.  This is an interesting subject.

Finally, we note that the original Tersoff parameters make the system melt into a liquid of wires while the ARK ones make the system melt into a 3D uniform liquid.
In view of the fact that the ARK parameters give a melting temperature closer  to the experimental one, we believe that they also describe better the melting of the silicene sheet.  The stability of a standalone
sheet of silicene at high temperatures however is not yet tested in experiments in spite of the fact that it is experimentally proved that silicene on Ag (111) surface is stable at room temperature.

{}

\end{document}